# Undulation-induced moiré superlattices with 1D polarization domains and flat bands in 2D bilayer semiconductors

Xingfu Li,* Sunny Gupta, and Boris I. Yakobson†

*Department of Materials Science and NanoEngineering, Rice University, Houston, TX, 77005 USA*

*Email: Xingfu.Li@rice.edu

†Email: biy@rice.edu

## ABSTRACT

Two-dimensional (2D) sheets typically have a high Föppl–von Kármán number, that is in-plane extent much greater than thickness, and can be easily bent, much like a paper, making such deformation a novel feasible way to design distinct electronic phases. Through first-principles calculations, we reveal the formation of 1D polarization domains and 1D flat electronic bands by bending-undulation of a 2D bilayer semiconductor. For a 1D sinusoidal deformation of a hexagonal boron nitride (*h*BN) bilayer as an example, we demonstrate how undulation induces nonuniform shear patterns, creating regions with unique local stacking and vertical polarization akin to sliding-induced ferroelectrics observed in twisted moiré systems. A particular sliding-induced polarization pattern also appears in double-wall BN nanotubes due to curvature differences between inner and outer tubes. Furthermore, undulation generates a shear-induced 1D moiré pattern that perturbs electronic states, confining them into 1D quantum well-like bands with kinetic energy quenched in modulation direction while dispersive in other directions (1D flat bands). This electronic confinement is attributed to modulated shear deformation potential resulting from tangential polarization due to the moiré pattern. Thus, bending modulation and interlayer shear it causes offer an alternative avenue, termed "curvatronics", to induce exotic phenomena in 2D bilayer materials.

## Introduction

Van der Waals structures [1], consisting of two-dimensional (2D) materials stacks with precise control over monolayer elements, thickness, and interlayer alignment, offers a unique platform for discovering exotic physical effects not present in the individual counterparts. One such system, twisted bilayers with moiré patterns arising from a small mismatch in the layers' periodicity have led to several paradigm-shifting physical phenomena, referred as twistronics [2]. The long-wavelength moiré period results in electronic flat bands, facilitating the study of strongly-correlated physics [3,4]. Additionally, twisting materials with intrinsic broken inversion symmetry, ranging from transition metal dichalcogenides to hexagonal boron nitride (*h*BN) and beyond [5], creates domains with permanent and switchable electric polarization, an effect coined slidetronics [6], with potential in ultrathin nonvolatile memory and computing devices. With twisted bilayer systems being extensively explored, we probe a distinct perturbation form, namely, bilayers that are periodically bent. Periodically bent bilayers are also expected to exhibit moiré patterns and artificial lattices due to changes in interlayer shear and local stacking. However, unlike twist-produced 2D moiré, uniaxial periodic bending with zero Gaussian curvature can generate 1D moiré patterns, leading to artificial electronic states, like 1D quantum wells and 1D ferroelectric domains, unattainable by twisting [7].

Bending 2D monolayers with non-zero Gaussian curvature has been an endeavor to create non-uniform strain pattern for generating pseudo-magnetic field [8–15], altering electronic structure and triggering exotic phases [16–18], such as Landau-level-like flat bands [9,10,13], correlated electronic states [10,14], valley Hall effect [11,12] and so on. Whether bending is achieved by suspending samples [19], loading them onto substrates [20], or using profiled substrates [11,15,21], an intralayer strain field is induced by local changes of bond lengths in the resultant artificial flexural lattices, raising potential mechanical stability issues of retaining such intralayer strains. This raises the question of whether artificial electronic states and phases can appear from pure shear in the absence of a modification to bond length, i.e., without intralayer strain. The Föppl–von Kármán number, proportional to the ratio of in-plane stiffness and out-of-plane bending rigidity, for typical 2D samples is $\gg 1$ [22], similar to a paper, signifying that 2D materials can easily bend without stretch or shrink. Bending bilayer, with no Gaussian curvature, causes an interlayer shear gradient, where the novel effects on electronic behavior have to be explored.

Here, by first-principles calculations, we explore the electronic properties of a periodically bent 2D bilayer with interlayer shear strain but no intralayer stress. Using an undulated *h*BN bilayer, a prototype 2D semiconductor as an example, we demonstrate that 1D undulation creates 1D moiré patterns due to variations in local stacking from the interlayer shear gradient. These undulations may form regions with different vertical electric polarizations, resembling 1D ferroelectric domains. Similar polarization domains also appear in BN nanotubes (BNNT), where stacking changes along the circumference lead to alternating electric polarization. Besides 1D polarization domains, the 1D moiré pattern induces long-range potential and 1D electronic confinement, creating 1D bands – flat in the modulated direction but dispersive in the unmodulated one, exhibiting peculiar van Hove singularities in the density of states (DOS). These undulated 2D systems, akin to 1D quantum wells, are easier to create by mechanically bending bilayers, without intricate heterogeneous synthesis. This work opens new avenues to create novel electronic phases such as 1D ferroelectric domains and 1D quantum-well-like electronic states using 2D bilayers undulated by the curved topography of substrate, thereby going beyond what's achievable with twists. We refer to our approach as further exploration in curvatronics [23], analogous to twistronics and straintronics.

## Results and discussions

***Undulation-induced slidetronics* -** A flat *h*BN bilayer can have *R*-type or *H*-type stacking based on the angle ($\varphi$) between its layers ($\varphi = 0°$ for *R*-type, and $\varphi = 60°$ for *H*-type). In the *R*-type (eclipsed) stacking, where B (N) lies directly above B (N), inversion symmetry is broken; sliding the layers also removes the out-of-plane mirror symmetry, producing an out-of-plane electric polarization arising from net charge transfer caused by the different electron affinities of B and N [6]. Such an out-of-plane electric polarization establishes an electrostatic potential difference in the layer-normal direction, hereafter referred to as the *potential step*. In contrast, *H*-type stacking maintains inversion symmetry and does not exhibit any out-of-plane polarization, regardless of sliding. Fig. 1a presents a 2D map of the sliding-induced potential step $\Delta V_{sl}$ of an *R*-type flat *h*BN bilayer as one layer is translated, consistent with the result in Ref. [24]. The distinct stacking configurations are indexed by the sliding ratios ($\nu_{zz}$, $\nu_{ac}$), defined as $\nu_{zz} = x_{zz}/c_{zz}$ and $\nu_{ac} = x_{ac}/c_{ac}$, where $x_{zz}$ ($x_{ac}$) is the displacement along zigzag (armchair) direction and $c_{zz}$ ($c_{ac}$) is the corresponding lattice constant of the rectangular unit cell (Fig. 1a inset). The *R*-type stacking serves as the reference (0, 0). Specifically, AB (1/3, 0) and BA (2/3, 0) stackings (see Fig. 1b inset) exhibit the largest and opposite potential steps. This out-of-plane polarization can be toggled by sliding between stacking geometries, with implications for ferroelectric applications [25,26]. Similarly, undulating a 2D bilayer changes interlayer shear, leading to different local stackings, and thus a similar out-of-plane polarization is expected in undulated *R*-type *h*BN bilayers.

To explore the aforementioned effect, we consider *h*BN bilayer undulated into sinusoidal shapes $f(x) = h\sin(2\pi x/L)$, with different amplitudes $h$ and period-lengths $L$. The atomic structures are designed to follow parallel curved surfaces with constant interlayer spacing and fixed bond length, as shown in Fig. 1c. The details of constructing geometry are provided as SM-1 in Supplemental Material [27]. In these structures, atoms on the left side of the supercell maintain the initial stacking, while interlayer shear increases and then decreases toward the right side, where atoms return to the initial stacking. The distinct shear patterns, characterized by the sliding ratio profiles throughout the undulated bilayer, for two amplitudes, $h = 3$Å and 6Å, at the same $L$, are shown in Fig. 1d.

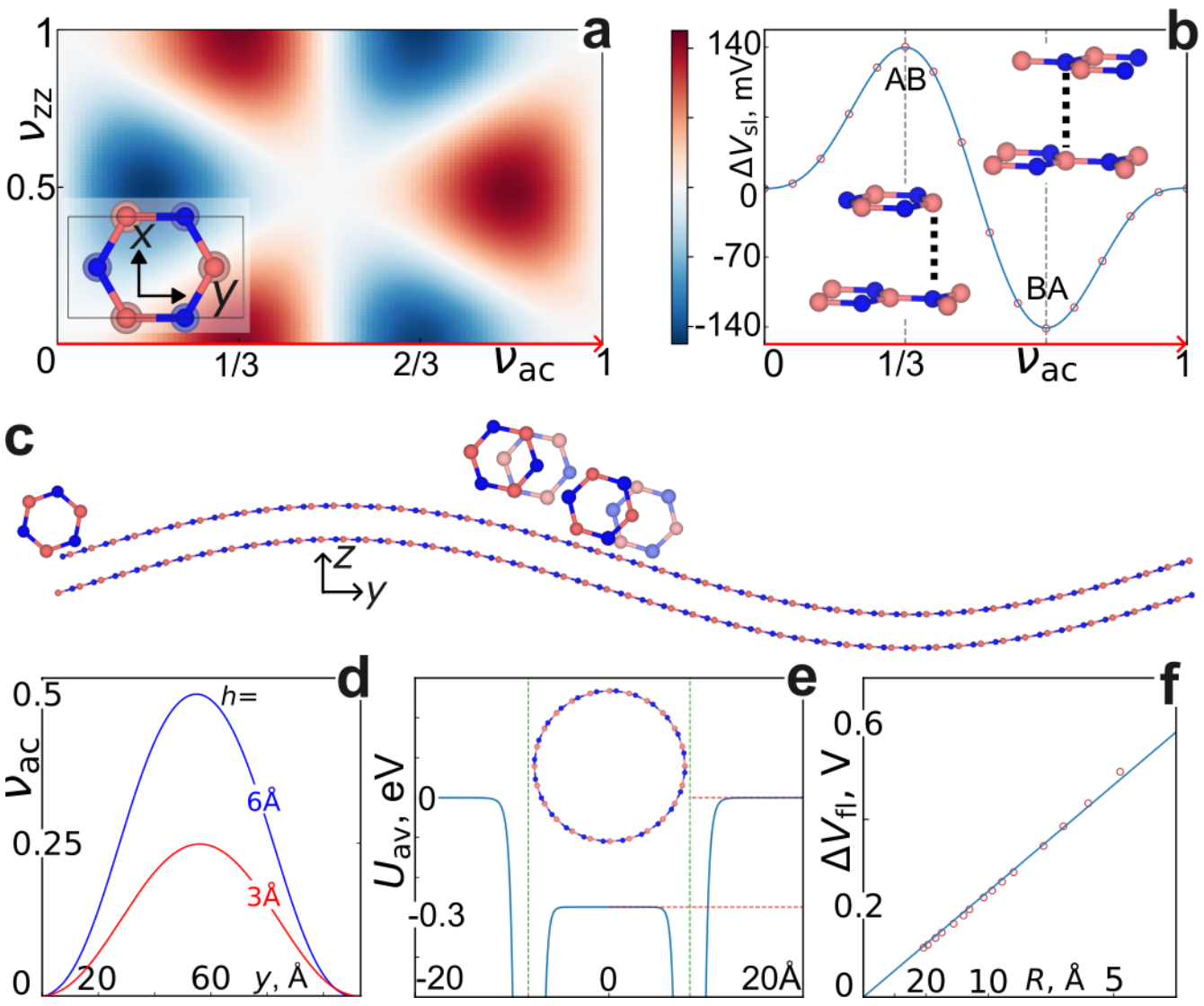

**Figure 1.** (a) A 2D map of the sliding-induced potential step $\Delta V_{sl}(\nu_{zz}, \nu_{ac})$. Inset shows the used unit cell of *R*-type flat *h*BN bilayer. Different stacking configurations are created by translational sliding between two layers. (b) The sliding-induced potential step of only armchair sliding $\Delta V_{sl}$ ($\nu_{zz} = 0$, $\nu_{ac}$). Insets illustrate the lattices of AB and BA stacked *h*BN bilayers, reddish for boron and blue for nitrogen. (c) Side view of a *R-ac-sin* bilayer. (d) Sliding ratios throughout the *R-ac-sin* bilayer for two bending

degrees: $h$ = 3 Å and 6 Å at $L_0$ = 113.1 Å ($L_0$ refers to the initial periodic length of flat bilayer before undulation as noted in SM-1 [27]). (e) The average electrostatic potential $U_{av}$ along the radial direction of a SW-BNNT. Green dashed lines indicate the diameter of the tube. (f) The linear relationship between the flexoelectric potential step $\Delta V_{fl}$ and tube curvature $1/R$.

In flat $R$-stacked $h$BN bilayer, sliding the layers along the armchair direction induces an out-of-plane polarization (Fig. 1a). We thus created an atomic structure of $R$-stacked $h$BN bilayer with sinusoidal modulation along the armchair direction (hereafter named *R-ac-sin* bilayer) (Fig. 1c). The local stacking in the undulated structure changes from $R$-type in the two ends to AB and BA in the middle (Fig. 1c), due to shear gradient, and should create out-of-plane polarization (Fig. 1a,b). Undulated bilayers are distinct from the flat, as out-of-plane polarization has two contributions: from different stacking $\Delta V_{sl}$, and flexoelectric polarization $\Delta V_{fl}$ from curvature [28–30]. Thus, the local total potential step across the undulated bilayer is $\Delta V_n(y) = \Delta V_{fl}(y) + \Delta V_{sl}(y)$, where $n$ indicates out-of-plane normal direction, and $y$ is a position in undulation direction. $\Delta V_{fl} = V_0/R$ is proportional to the local curvature $1/R$ ($V_0$ is the flexoelectric constant).

We obtain $V_0$ = 2.32 VÅ by calculating $\Delta V_{fl}$ for single wall boron nitride nanotubes (SW-BNNT) of different radii (Fig. 1d) using density functional theory (DFT). In Fig. 2a, the red and green dots represent the across-plane flexoelectric potential step, $\Delta V_{fl}(y)$, extracted from DFT calculation for two isolated single layers (which together form the bilayer). The analytical estimate, given by $\Delta V_{fl}(y) = V_0/R(y)$ and depicted as solid green and red curves labeled *flexo*, shows excellent agreement with the DFT results. In bilayers, a local stacking configuration S depends on relative translational interlayer shift, and can be mapped to an out-of-plane polarization $S \rightarrow P$ [24], so we obtain $\Delta V_{sl}(y)$ as the potential step measured in the identically stacked flat configurations (Fig. 1a). DFT-computed local total potential step $\Delta V_n(y)$ in undulated bilayers (blue dots in Fig. 2a) align well with the blue solid line representing the analytical estimate $\Delta V_n(y) = 2V_0/R(y) + \Delta V_{sl}(y)$. To further isolate, for an undulated bilayer, the out-of-plane polarization domains and potential induced by stacking change, we subtract the potential from the individual layers (I and II) from the total, $\Delta V_{sl}(y) = \Delta V_n(y) - \Delta V_{fl}(y)_I - \Delta V_{fl}(y)_{II}$. (See SM-2 [27] for details on how to extract local potential steps.) The undulation-induced $\Delta V_{sl}(y)$ is depicted in pink in Fig. 2a. The data obtained from DFT calculations are shown as circles, while the stacking-based estimates are shown as the solid curve. The alternating sign of potential steps confirm out-of-plane polarization domains in undulated bilayers, similar to the ferroelectric domains discovered recently in twisted-$h$BN [25,26]. The difference being that the domains in the present case arise due to shear, and are 1D. The domains' polarization can be switched by similar means as that obtained in the twisted-case.

It is also non-trivial to compare the relative magnitude of $\Delta V_{sl}$ and $\Delta V_{fl}$ achievable in a given *R-ac-sin* bilayer. Since $\Delta V_{sl}$ and $\Delta V_{fl}$ can be approximated by local stacking S and local curvature $1/R$, respectively, their maximum values are evaluated as functions of the amplitude $h$ and periodic length $L$. It is found that $\Delta V_{sl}$ generally dominates except when the bending is small, $h/L \rightarrow 0$, and it saturates when AB or BA stacking is locally achieved (Fig. S6 in Supplemental Material [27]).

It should be noted that in-plane polarization also develops in an undulated $R$-stacked $h$BN bilayer; a more comprehensive polarization distribution is provided in Fig. S7. Finally, only undulated $R$-stacked $h$BN bilayer, being not-centrosymmetric, exhibits out-of-plane polarization domains due to change in local stacking. The undulated $H$-stacked bilayer has local inversion symmetry, and the total out-of-plane potential step, as shown in Fig. S8, is only contributed from flexoelectric effect $\Delta V_n(y) = \Delta V_{fl}(y)_I + \Delta V_{fl}(y)_{II}$.

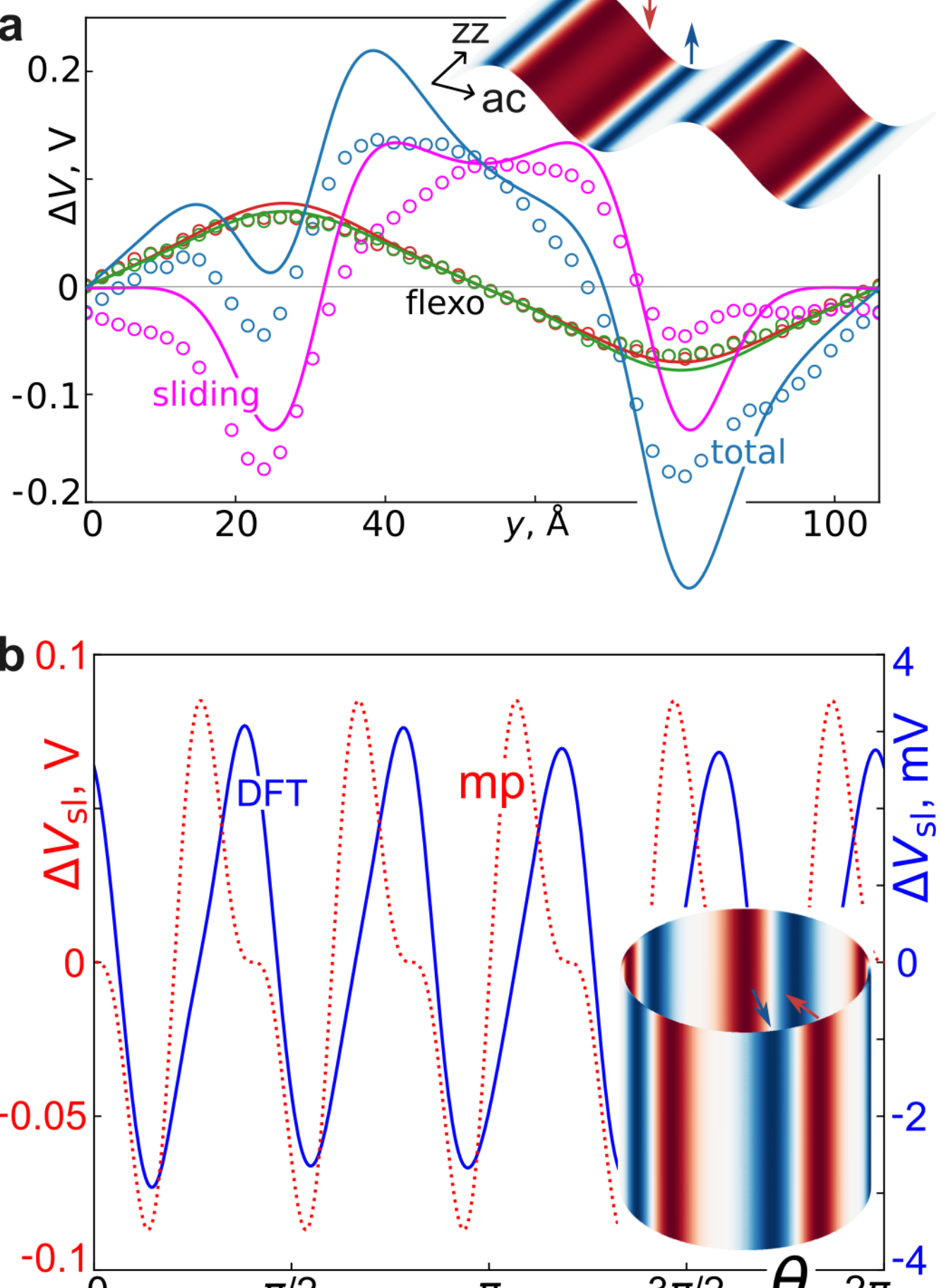

**Figure 2. (a)** Different components of the local potential step across an *R-ac-sin* bilayer with $h = 9$ Å and $L_0 = 113.1$ Å. The circles and solid curves represent data from DFT calculation and local mapping, respectively. The red and green colors refer to the local potential step purely due to bending flexoelectricity in two individual sinusoidal monolayers, of which analytical fitting is derived from the linear relationship between local curvature and flexoelectric potential step. The pink color refers to the local potential step purely due to sliding. The blue color refers to the local total potential step across the bilayer, of which analytical fitting is the addition of all components from analytical models. The colormap sketches the polarization domains of two periods in armchair (ac, $y$) direction due to sliding, where arrows indicate out-of-plane dipole direction, and $\Delta V_{sl}$ ranges from -0.14 V to 0.14 V. **(b)** The sliding-induced local potential step $\Delta V_{sl}(\theta)$ across the double nanotube walls for (7, 7)@(12, 12) acDW-

BNNT. The red curve is obtained by mapping (mp) local stacking to flat configuration, while the blue curve is directly extracted from DFT at a given radius-distance $r$. The colormap sketches the polarization domains in acDW-BNNT. Due to a larger intertube distance compared to the interlayer distance above [6], $\Delta V_{\mathrm{sl}}$ is reduced to lie between -0.08 V and 0.08 V.

Sliding-induced polarization domains should also occur in double wall boron nitride nanotubes (DW-BNNT), where the difference in the walls' curvature creates shear, altering local stacking. For simplicity, we focus on armchair DW-BNNT (acDW-BNNT) with chiral indices ($n$ - 5, $n$ - 5) and ($n$, $n$). In acDW-BNNT, shear occurs along the armchair direction, with local areas experiencing AB and BA stackings, which result in the maximum out-of-plane potential step. The periodicity of the shear pattern around the acDW-BNNT circumference depends on the radius difference between the tubes. For an acDW-BNNT with an intertube distance $d$ = 3.46 Å and bond length $l$ = 1.45Å, the total period in shear pattern is expected to be $2\pi d/(3l) \sim 5$, as a shear displacement by three bond lengths completes one interlayer sliding cycle. Notably, this means that the symmetry in the domain pattern, seen in Fig. 2b, and its inset is invariantly 5-fold, regardless of the nanotube diameter. Similar to undulated bilayers, the polarization domains in acDW-BNNT are mapped from local stacking configurations S. The red curve in Fig. 2b shows the alternating sliding-induced potential step $\Delta V_{\mathrm{sl}}(\theta)$ across the varying polarization domains, based on the local S→$P$ mapping from Fig. 1a.

To further confirm the presence of polarization domains, $\Delta V_{\mathrm{sl}}(r,\theta)$ from DFT calculations is plotted in the azimuthal direction $\theta$ at a radius-distance $r$ (> $r_{\text{outer tube}}$, Fig. S9), shown by the blue curve in Fig. 2b. Following the approach used for undulated bilayer $h$BN, we define $\Delta V_{\mathrm{sl}}(r,\theta) = \Delta V_{\mathrm{DWNT}}(r,\theta) - \Delta V_{\mathrm{inner}}(r,\theta) - \Delta V_{\mathrm{outer}}(r,\theta)$, where $\Delta V(r,\theta) = V(r,\theta) - V_{\mathrm{axial}}$ in respective calculation. Since the flexoelectric effect is uniform at a fixed $r$ and does not depend on $\theta$, $\Delta V_{\mathrm{inner}}(r,\theta)$ and $\Delta V_{\mathrm{outer}}(r,\theta)$, are constants. The potential profile $\Delta V_{\mathrm{sl}}(r,\theta)$ has five periods, consistent with the expected shear pattern. Additionally, the agreement between the behavior of the potential profile obtained from DFT calculations and mapping from local S → $P$, confirms bending-induced shear and existence of out-of-plane polarization domains in acDW-BNNT. Our examples of undulated $R$-stacked $h$BN bilayer and acDW-BNNT establish the functional form of sliding-induced 1D out-of-plane polarization domains, a result of shear-induced modulation of local stackings.

***Undulation-induced 1D quantum-well-like states:*** Concurrent to forming ferroelectric domains, 1D undulation necessarily generates a broader moiré potential via stacking alterations, similar to twisted-moiré systems. Such a moiré potential is ubiquitous and is expected to confine electrons along the undulation direction in any generic undulated 2D bilayer semiconductor. To illustrate the effect of undulation on the electronic structure, we employ sinusoidal $h$BN bilayers, which serves as a representative model for a wide range of 2D semiconductors. Although a sliding-induced potential also exhibits a moiré pattern, it is not considered in the subsequent analysis. In particular, we consider an $H$- stacked *h*BN bilayer, lacking sliding-induced out-of-plane polarization, to exclude the effect of sliding-induced potential on the electronic structure. Specifically, we investigate the electronic band structure along and perpendicular to the undulation direction, corresponding to the **k**-path ($\pi/L$,0,0)-(0,0,0)-(0,$\pi/L$,0), noted as X-Γ-Y, in an *H*-stacked *h*BN bilayer that is sinusoidally modulated along the zigzag direction (hereafter referred to as the *H-zz-sin* bilayer), with the modulation oriented along the *x* direction in real space by design. After modulation, the initially parabolic folded bands along X-Γ (Fig. S10) at band edges (near valence band maximum, VBM, and conduction band minimum, CBM), in flat *H*-stacked bilayer, become flattened and isolated, while bands along Γ-Y remain dispersive (Fig. 3a). As bending increases, the evolution of band edges from dispersive to flat along X-Γ is clearly seen (Fig. S11).

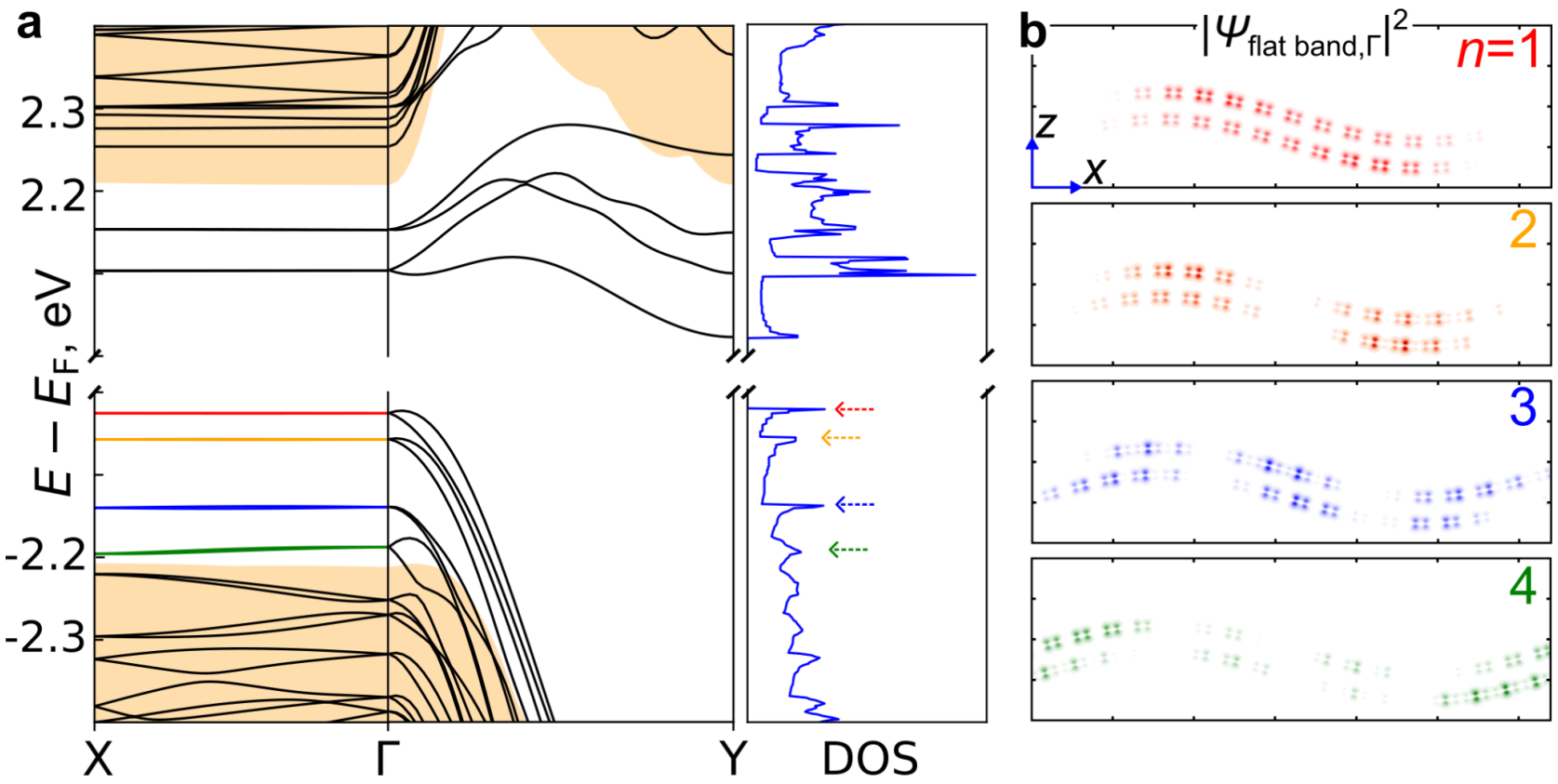

**Figure 3. (a)** Electronic band structure and density of states (DOS) from DFT calculation of *H-zz-sin* bilayer with $h$ = 3 Å and $L_0$ = 65.3 Å. Yellow-shaded areas mark the bands of flat *H*-stacked bilayer. The length of **k**-path X-Γ is rescaled for visual clarity due to the large lattice constant of modulation direction. The van Hove singularities induced by 1D states in flat valence bands are indicated by arrows. **(b)** The spatial distributions of electronic states, $|\Psi_{\text{flat band},\Gamma}|^2$, corresponding to the Γ point of 1D flat valence bands. All shown states are indexed ($n$ = 1-4) by the node feature of quantum states of the 1D quantum well model. The colors of indices show correspondence to flat bands in (a).

To further analyze the nature of 1D flat bands, we examine the spatial distributions of electronic states in flat bands at Γ point, $|\Psi_{\text{flat band},\Gamma}|^2$. The highest 1D flat valence (double degenerate) bands (Fig. 3a red bands) are spatially-localized at one domain (Fig. 3b topmost panel, $n$ = 1 state); that of the lower 1D flat valence bands (Fig. 3a orange bands) is localized at two domains (Fig. 3b next panel, $n$ = 2 state). The states $n$ = 3, 4 are also shown. Similar localization also exists (Fig. S12) for flat conduction bands. We note that, for each flat band, the distribution of state at every $k$ between X and Γ is identical to that at Γ and, in contrast, there is no localization for the distribution along $y$ direction (Fig. S22). These localized states in a series of 1D flat bands resemble the eigenfunctions of a 1D quantum well model (1D model). Generally, states with more nodes have higher energy, however, here, the 1D flat valence bands have hole character, where the direction of increasing energy is opposite of that for electrons. The hole states with more nodes and greater dispersion appear at lower electron energy. The presence of 1D flat bands and its close resemblance with a 1D model implies the presence of a 1D electronic confinement potential throughout the undulated bilayer, arising possibly due to the moiré pattern. In a monolayer undulated in the same way, no such 1D flat bands or localized states are found near the VBM and CBM (Fig. S13), indicating that the interlayer coupling in bilayer is crucial to create the modified electronic states. In an undulated bilayer, varying interlayer shear changes the local stacking and can alter the interlayer coupling, affecting the local electronic energy. Therefore, the 1D confinement moiré potential responsible for the 1D flat bands could originate from the deformation of the electronic states due to changing shear, much like the commonly studied deformation caused by tensile or compressive strain [31].

To elucidate the origin of the moiré confinement potential, we evaluate the contribution of shear ($\gamma$) deformation potential, which is defined as the variation in VBM relative to the vacuum level across flat bilayers with different stacking configurations. In this section, we consider only shear along the zigzag direction and denote the corresponding shear deformation potential as $U_\gamma(v_{zz})$. Consequently, the deformation potential profile throughout an undulated bilayer as a function of $x$, $U_\gamma(x)$, is determined by mapping the local stacking configuration, $v_{zz}(x)$, onto the VBM of flat bilayer with identical stacking. Fig. 4a shows $U_\gamma(x)$ for three *H-zz-sin* bilayers with various bending amplitudes. To highlight the resemblance with the 1D model, $U_\gamma(x)$ is used as an external potential to numerically solve the 1D Schrodinger equation, 1DSE: $[-(\hbar^2/2m_e)\nabla_x^2 + U_\gamma(x)]\psi = E\psi$. The resulting dispersions naturally show discrete energy levels due to confinement (Fig. S15), with energy separation agreeing well with the eigenenergy difference between 1D flat valence bands obtained from DFT calculations (Fig. S16 and Fig. 3a). This confirms the similarity, in terms of confinement to electrons and holes, between undulated bilayer and the 1D model, as well as that the shear deformation potential likely creates the electronic confinement.

Probed this way, shear deformation potential can be used in 1D model to quantify how the band edges evolve, from dispersive to flat, when the flat-bilayer gradually bends, i.e., in the limit $h/L \rightarrow 0$. Two quantities describing the band dispersion are of particular interest: the band width $W$, which is the energy range of the topmost band in the spectrum, and the energy separation $E_s$ from the highest flat band ($n$ = 1) to the nearby continuum. In the context of DFT calculation, $E_s$ represents the energy difference between the VBM in undulated bilayer and that in flat bilayer (Fig. S17). Both $W$ and $E_s$ depend on the confinement potential, determined by undulation parameters $h$ and $L$. In the limit $h/L \rightarrow 0$, the position-dependent deformation potential profile exhibits a quadratic dependence on $h/L$, $U_\gamma(h/L,x) = -145(1 - \cos(2\pi x/L))^2 h^2/L^2 = -290\sin^2(\pi x/L)\cdot h^2/L^2 = u_\gamma(x)h^2/L^2$ (detailed derivation in SM-3 [27]). This quadratic behavior arises because $U_\gamma$ is symmetric and independent of the bending direction, which justifies the quadratic approximation $U_\gamma(v_{zz}) = -2.19v_{zz}^2$, valid for small sliding ratio $v_{zz}$ (Fig. S14). Additionally, $v_{zz}$ is linearly related to $h/L$, confirming the quadratic relationship between $U_\gamma(x)$ and $h/L$.

Upon determining, $U_\gamma(x) \propto h^2/L^2$, the behaviors of $W(h, L)$ and $E_s(h, L)$ can be further derived. Using the eigenvalue solutions of Mathieu's equation [32,33], an analytical form of $W(h, L) = (c_1 + c_2h^2)/L^2$ is obtained (details in SM-4 [27]), where $c_{1,2}$ are fitting constants. Additionally, $E_s(h, L)$ is determined by first-order perturbation theory, $E_s = \langle \psi^{(0)} | U_\gamma | \psi^{(0)} \rangle = \dfrac{h^2}{L^2} \langle \psi^{(0)} | u_\gamma(x) | \psi^{(0)} \rangle$, where $\psi^{(0)}$ is the eigenstate of interest, yielding $E_s(h, L) = c_3(h/L)^2$, with $c_3$ as a fitting constant. Fig. 4e shows the dependence of the band width $W$ (orange surface) and the separation from the continuum $E_s$ (blue surface) on $h$ and $L$. The analytical results $W(h, L)$ and $E_s(h, L)$ are fitted to their values obtained by numerically solving the 1DSE for various combinations of $h$ and $L$ (Fig. S18), in good agreement. It can be beneficial to discuss $W(h, L)$ behavior in three scenarios. First, as $h$ increases with a fixed $L$ (red line in Fig. 4e), $W$ decreases due to stronger localization by a deeper confinement potential. Second, when $h/L$ is fixed with increasing $h$ or $L$ (green line), the potential depth does not change, the decrease of $W$ is attributed to the larger separation between neighboring localized states. Third, with fixed $h$ and increasing $L$ (blue line), $W$ tends to decrease, largely for the same reason as in the second case. The analytical equations of $W(h, L)$ and $E_s(h, L)$ provide a predictive framework for understanding the evolution of band edges as the bilayer bends.

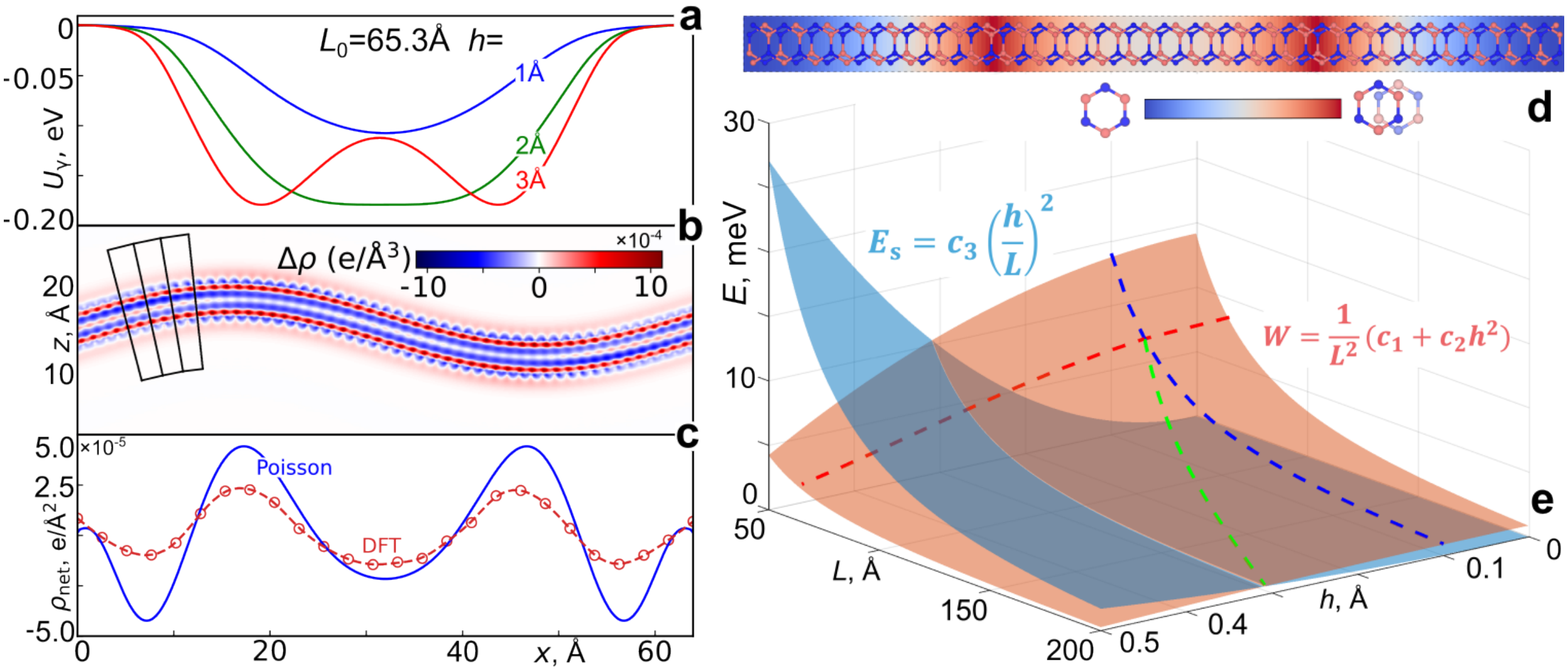

**Figure 4. (a)** Shear deformation potential profile $U_\gamma$ as a function of position in undulation direction. Shown are the $U_\gamma$ values for *H-zz-sin* bilayers of three amplitudes $h$. **(b)** 2D map of differential charge density $\Delta\rho(x,z)$ for *H-zz-sin* bilayer with $h = 3$ Å and $L_0 = 65.3$ Å. **(c)** Comparison between $\rho_{net}$ extracted from DFT-calculated $\Delta\rho$ (red) and charge density distribution analytically solved from $U_\gamma(x)$ by Poisson's equation (blue). **(d)** Top view of the crystal structure with the local sliding ratio overlaid as a colormap, revealing the moiré pattern. Note that in the current geometry, local stacking is resolved only from a view normal to the local surface, so the moiré pattern cannot be discerned from a single top view. **(e)** The dependence of band width $W$ (orange surface) and energy separation from continuum $E_s$ (blue surface) on $h$ and $L$, plotted according to the analytical expressions.

It should be noted that the *H-zz-sin* bilayer shows no out-of-plane polarization. Therefore, the electronic confinement arises solely from a modulating shear deformation potential along the tangential direction, due to the varying sliding ratio in the 1D moiré stacking pattern (Fig. 4d). Such electrostatic potential must originate from charge redistribution, a tangential polarization. To verify charge redistribution along the tangential direction in the undulated bilayer, we calculate the differential charge density $\Delta\rho = \rho_{\mathrm{bilayer}} - \rho_{\mathrm{I}} - \rho_{\mathrm{II}}$, where $\rho_{\mathrm{I}}$ and $\rho_{\mathrm{II}}$ are the electronic charge densities of individual layers I and II, and visualize it as a 2D map in the $xz$ plane, with the $y$ distribution averaged (Fig. 4b). We then employ a macroscopic averaging technique, tailored for curved, layered atomic structures, to obtain the local net charge density, $\Delta\rho_{\mathrm{ave}}$, in order to better reveal electrostatically non-neutral regions in *H-zz-sin* bilayer. In detail, the local net charge density is defined as $\Delta\rho_{\mathrm{ave}} = \int_v \Delta\rho \mathrm{d}v/V$, where $V$ represents a local volume for macroscopic averaging, depicted by the black box in Fig. 4b. The selection of $V$ ensures that its in-plane (parallel to atomic structure) dimensions match the $xy$ plane of the rectangular unit cell (Fig. 1a inset), and its out-of-plane (normal to atomic structure) dimension is sufficiently large to ensure convergence in the average charge density calculation (see Method for more details). Consequently, $\Delta\rho_{\mathrm{ave}}$ is a function of $x$ (the undulation direction), highlighting the 1D feature of the present model. A value of $\Delta\rho_{\mathrm{ave}}(x) = 0$ indicates that electroneutrality is maintained locally—as is the case in a flat bilayer configuration—whereas any deviation from zero signifies charge flow along the tangential direction.

To give a more intuitive physical quantity, we define the local net charge density per area as $\rho_{\mathrm{net}} = \Delta\rho_{\mathrm{ave}}/(ab)$, where $a$ and $b$ are lattice constants of the rectangular unit cell (Fig. 1a) for zigzag and armchair direction, respectively. Red curve in Fig. 4c shows $\rho_{\mathrm{net}}(x)$ of different regions throughout the undulated bilayer, where 26 unique values are obtained, corresponding to 26 rectangular unit cells in building the supercell of sinusoidal structure. A negative and positive net charge density indicates a negatively and positively charged local area, respectively. Therefore, the charge redistribution along the tangential direction is verified directly from the results of DFT calculation. The $\rho_{\mathrm{net}}(x)$ is also compared with the charge density distribution corresponding to the shear deformation potential $U_\gamma(x)$, solved from 1D Poisson's equation $\nabla^2 U_\gamma(x) = \rho_{\mathrm{net}}(x)/\varepsilon_0$ (blue curve in Fig. 4c), and the trends are in agreement. Accordingly, the shear-induced deformation potential through the undulated bilayer is proved to be the main cause of the 1D flat bands. To convey the idea better yet, a simple tight-binding model is constructed in SM-5 [27] to show that 1D confining potential can unidirectionally flatten electronic bands even in a flat $h$BN monolayer. In the present undulated bilayers, shear deformation potential serves as the cause of confinement. Further, the resulting 1D states are shown to be robust even if the deformation potential is moderately perturbed by the defects in 2D layers or disorder in supporting substrates.

We emphasize that charge redistribution in the tangential direction, which creates the shear deformation potential, leads to the presence of in-plane dipole moments. In *H*-stacked $h$BN bilayer with inversion symmetry, no polarization in the uniform flat configuration, unlike the intrinsic in-plane 2D ferroelectrics [34], this tangential polarization from the 1D moiré pattern drives the electronic confinement and 1D flat bands. However, this in-plane polarization has been rarely discussed in the recent slidetronics works of 2D moiré systems, and the condition of *H*-stacked bilayer is completely overlooked [24]. It could be insightful to investigate the detailed polarization texture in the moiré-based configurations in the future.

The 1D quantum-well-like flat bands along X-Γ are also present in *R*-stacked *h*BN bilayer sinusoidally modulated in the zigzag direction (*R-zz-sin* bilayer). Although the distribution of $|\Psi_{\mathrm{flatband},\Gamma}|^2$ is less regular, but the quantum-well-like node character is still observable (Fig. S19), indicating that the origin of these 1D bands does not depend on the specific stacking. It is worth mentioning that sinusoidal modulation along armchair direction in either *H*- or *R*- stacked *h*BN bilayer (*H-ac-sin* and *R-ac-sin* bilayer) can also create similar 1D flat bands along Γ-Y, rather than X-Γ (Fig. S20). However, the electronic bands along Γ-Y in such systems reside at deep energy levels, which are far below (above) the VBM (CBM), thus rendering them not of much interest. Hence, only *h*BN bilayers undulated along the zigzag direction can create 1D quantum-well-like states near the valence and conduction band. These 1D states display van Hove singularities, which if accessed upon electron/hole doping, can be a novel physical system to realize strongly correlated physics phenomena, such as superconductivity, magnetism, strongly correlated electronic phases [35–37], etc. Additionally, the 1D feature may offer a way to mimic the multiband materials with quasi-1D band in the context of superconductivity [16,38–40] and electronic anisotropy [11,41], and study the interplay between quasi-1D states and states with high dimensional characteristics [42]. On the other hand, these 1D bands are analogous to those in widely studied 1D quantum wells based on bulk III-V semiconductors [43]. Similar to 1D quantum wells, these 1D states in undulated 2D bilayers can greatly enhance the lifetime of photogenerated carriers and increase the overall performance of devices based on such effects, which are beneficial for applications in quantum cascade lasers, enhanced photogenerated carrier efficiency for photovoltaics, LEDs, and photodiodes (see SM-6 [27] for more discussions on possible applications).

To realize the predicted effects, bilayer *h*BN can be overlaid, stamped, or potentially grown directly on patterned substrates—such as sinusoidally structured $SiO_2$—already demonstrated via lithographic techniques having aspect ratios >0.3 [44]. Moreover, periodic sinusoidal ripples with aspect ratios up to ~0.08 have been achieved in graphene sheets suspended over Si/$SiO_2$ trenches [45], and similar wrinkle patterns form in 2D materials on relaxed pre-strained polymers [46,47], suggesting the strategies feasible for bilayer *h*BN.

## Conclusions

In conclusion, we have shown that the undulation, in one direction, of a 2D van der Waals bilayer can generate a 1D moiré pattern, inducing sliding-induced ferroelectric domains paired with flexoelectric polarization. The idea is also extended to DW-BNNT, where the texture of polarization domains in acDW-BNNT is discussed. The sliding-induced polarization domains are similar to those in recently discovered twisted-moiré systems and can be switched to demonstrate 1D ferroelectric behavior. Additionally, the out-of-plane polarization due to undulations is an outcome of a combined effect of sliding-induced ferroelectricity and curvature-induced flexoelectricity, which increases the magnitude of surface potential modulation on either side (top/bottom) of the layers. We find that a total modulation of $\Delta V_n \sim 0.5$ eV (Fig. 2a) can be achieved in such undulated layers with moderate aspect ratios $h/L < 0.1$. This large $V_n$ can alter the band positions of another functional layer like $MoS_2$, when stacked on top of the undulated *h*BN bilayer and can significantly impede exciton diffusion [48] and modulate the optical properties of the functional layer.

On the other hand, by choosing a specific modulation direction, the emergence of 1D flat bands due to 1D moiré confinement potential in undulated *h*BN bilayer is shown at the top/bottom of valence/conduction bands. The confinement potential is attributed to shear deformation potential modulation arising from the tangential polarization due to the moiré pattern. Such in-plane polarization has been overlooked before in moiré-based systems until a recent study exclusively focused on flat *R*-stacked *h*BN bilayer [24], and must be included to get accurate representation of the polarization textures. The 1D confinement potential in undulated *h*BN bilayer itself is also expected to confine excitons and realize highly anisotropic exciton flow [49]. The 1D quantum-well-like flat bands display van Hove singularities, which if accessed upon proper electron/hole doping can be employed to study strongly correlated physics in 1D. Moreover, due to the similarity between localized states in 1D flat bands and those in conventional quantum wells, these states can be exploited to design related high-performance functional devices, such as quantum cascade laser, photovoltaics, LEDs and more [50,51]. *h*BN is one prototype example, we expect similar undulation-induced moiré superlattices with 1D polarization domains and flat bands in other 2D semiconductors, such as transition metal dichalcogenides [52].

## Methods

***First-principles calculations:*** Density functional theory (DFT) calculations were performed using DFT-D2 van der Waals (vdW) approach [53] in conjunction with Perdew−Burke−Ernzerhof (PBE) generalized gradient approximation [54] in the framework of all-electron projector augmented wave (PAW) method [55] as implemented in the Vienna Ab-initio Simulation Package [56]. The plane-wave kinetic energy cutoff was 500 eV. To eliminate the interaction between the bilayer with its periodic images, the supercell lattice constant in *z* direction had been set as 30 Å. For *R*- or *H-zz* (*R*- or *H-ac*) bilayer, *k*-mesh of 1×20×1 (20×1×1) was used in self-consistent field calculation until the total energy was converged to $10^{-5}$ eV.

The sinusoidal monolayer was first generated, and the supercell lattice constant along undulation direction was chosen so that the arc length of one sine period equaled the original lattice constant of flat configuration. (The details of constructing geometry are provided in SM-1 [27].) Then, ionic relaxation with fixed cell boundary (to keep the designed geometry) is performed using DFT until the force on each atom falls below 0.02 eV/Å. The atom positions almost do not change during the ionic relaxation, which is expected since no intralayer strain is introduced upon assigning sinusoidal modulation.

The sinusoidal bilayer was generated based on the idea of parallel curves. The ionic relaxation of the obtained bilayer (using DFT, with fixed cell boundary) did not show any faceting and kinks except for up to 6% interlayer distance change due to different vdW correction choice. This is because our shape is small, thus it maintains quite precisely its designed sinusoidal shape. A supplementary energy analysis about faceting is provided in SM-1 [27]. The parallel curves method indeed appropriately captures the variation in interlayer shear inherent to the sinusoidal bilayer. The structure obtained from the parallel curves method is employed for electronic band structure calculations. The **k**-path ($\pi/L$,0,0)-(0,0,0)-(0,$\pi/L$,0), noted as X-Γ-Y, in the Brillouin zone is used, which has different absolute values for supercells with different sizes.

***Macroscopic averaging:*** Macroscopic averaging is performed upon a 2D grid data set of charge density (encoded in $x$ and $z$ coordinates, for a *H-zz-sin* bilayer), which is derived by averaging the raw volumetric data along $y$ direction (the direction without undulation in a *H-zz-sin* bilayer). The idea is to partition the central line between two layers (two curves in the projected $xz$ plane) into 26 pieces, corresponding to the 26 rectangular unit cells in building the supercell of sinusoidal structure. The length of each segment equals the lattice constant of zigzag direction in the unit cell. The tangential size of averaging window, black box in Fig. 4b, follows these segments and the two longer sides are strictly normal to the central line and the two layers. Therefore, the averaging window has a shape of rectangle when the local curvature of the atomic structure is zero, such as left/right end and mid, otherwise, a shape of trapezoid, while the size of the averaging window is fixed. In each averaging window, average charge density difference can be calculated by $\Delta\rho\text{ave} = \int v \Delta\rho dv/V$.

## Acknowledgements

Funding resources.
The authors thank Jun-Jie Zhang and Yuefei Huang for insightful discussion and advice.